\documentclass[12pt]{article}

\usepackage{authblk}
\usepackage{graphicx, setspace, amsmath, amsfonts, amssymb, fontenc,  rotating,}
\usepackage{makecell, gensymb, secdot}

\usepackage[top=1in, bottom=1in, left=0.7in, right=0.7in]{geometry}
\usepackage[utf8]{inputenc}
\usepackage{pdflscape}
\def\bp{\begin{pmatrix}}
\def\ep{\end{pmatrix}}
\def\bc{\begin{center}}
\def\ec{\end{center}}
\def\be{\begin{equation}}
\def\ee{\end{equation}}

\begin{document}
\title{Hierarchy of CKM matrix elements and implications of unitarity}

\author{Gurjit Kaur, Gulsheen Ahuja$^*$, Dheeraj Shukla and Manmohan Gupta$^\dagger$\\Department of Physics,\\ Panjab University, Chandigarh, India.\\
\vspace{0.7cm}
* gulsheen@pu.ac.in\\
$^\dagger$ mmgupta@pu.ac.in}

\onehalfspacing
\maketitle

\begin{abstract}
The hierarchy amongst the CKM matrix elements, highlighted recently by Luo and Xing, has been rigorously revisited using the PDG parameterization incorporating unitarity constraints. Further, we have explored the evaluation of the CP violating parameter $\epsilon_k$ for the 9 possible unitarity ensured equivalent independent parametrizations of CKM matrix. Interestingly, we find that not all of these reproduce the value of parameter $\epsilon_k$, this being presumably due to the hierarchical nature of the CKM matrix elements. This situation is echoing similar conclusions regarding the evaluation of Jarskog's rephasing invariant parameter J through unitarity ensured equivalent possibilities.
\end{abstract}

\section{Introduction}
Over the years, both on the theoretical as well as  the experimental front, important developments have taken place regarding the  Cabibbo-Kobayashi-Maskawa  (CKM) \cite{ckm1,ckm2} phenomenology. The theoretical efforts have highlighted the crucial role of CKM paradigm  in understanding several features of flavor physics. On the experimental front, several  groups like Particle Data Group (PDG) \cite{pdg24}, CKMfitter \cite{ckmfit}, HFLAV \cite{hflav23}, UTfit \cite{utfit}, etc., have been actively updating their analyses to arrive at more and more refined data. These efforts have resulted in almost precise determination of various CKM parameters. Remarkable progress in the context of CKM phenomenology has given way to further exploring the consequences of unitarity having implications for precision measurements of CKM matrix elements. 

In this context, Xing \textit{et al.}, in a few earlier papers \cite{asy1, asy2} and in a recent one \cite{asy3} have explored some interesting constraints on the elements of the quark mixing matrix. In particular, using unitarity constraints, within the framework of  the modified Wolfenstein parametrization, the authors have arrived at an interesting hierarchy among the 9 CKM matrix elements. Further, they have made an attempt to determine the  Jarskog's rephasing invariant parameter J through 4 moduli of the 9 quark mixing matrix elements. In this context, the authors \cite{asy3} have enumerated large number of possibilities out of which most of these are not able to reproduce the value of the parameter J. These considerations provide a motivation for a re-visit of some aspects of the CKM phenomenology wherein the equivalent possibilities which are ensured by unitarity are not compatible with data.

It may be noted that recently \cite{cartesian} we have constructed 9 possible independent parametrizations of CKM matrix in rigorous and ab-initio manner, starting with each of the 9 elements of the matrix. It has been observed that out of the 9 representations, the PDG representation, wherein unitarity is in-built, unlike the Wolfenstein parametrization, perhaps provides the best option for carrying out CKM phenomenology at the present level of measurements. Therefore, it becomes interesting to rigorously re-visit the hierarchical relations, incorporating unitarity, among the CKM matrix elements through the PDG parametrization. Further, in order to check these hierarchical constraints numerically, we would reconstruct the magnitudes of the CKM matrix elements by using well known elements and by incorporating unitarity constraints. Also, it would be interesting to explore whether there are other situations in CKM phenomenology, similar to the example of evaluation of J \cite{asy3}, wherein out of all possible equivalent choices, only a few possibilities emerge out to be viable to reproduce parameter J. Interestingly, we find that while evaluating the CP violating parameter $\epsilon_k$ using the 9 possible equivalent independent parametrizations of CKM matrix, again, not all of these provide an appropriate fit to parameter $\epsilon_k$.

\section{Hierarchy of the CKM matrix elements}
To begin with, we define the CKM matrix as
 \be \begin{pmatrix} d' \\s'\\b' \end{pmatrix} = V_{CKM} \begin{pmatrix} d\\s\\b\end{pmatrix}, ~~~~~~\text{where}~ V_{CKM}=\begin{pmatrix} V_{ud}& V_{us}& V_{ub}\\ V_{cd}& V_{cs}& V_{cb}\\V_{td}& V_{ts}& V_{tb}\end{pmatrix}\ee
 is a unitary matrix. Unitarity of the CKM matrix is expressed in terms of the following  orthogonality and  normalization conditions of the CKM matrix
  \be{\displaystyle \sum _{\alpha}V_{i\alpha}V_{j\alpha}^{*}=\sum _{i}V_{i\alpha}V_{i\beta}^{*}=0~,}\label{ortho} \ee 
\be {\displaystyle \sum _{\alpha}|V_{i\alpha}|^{2}=\sum _{i}|V_{i\alpha}|^{2}=1}, \label{nor}\ee 
 respectively, where $i,~j\equiv (u,~c,~t)$ and $\alpha, ~\beta\equiv(d,~s,~b)$. 
 
 In order to have a re-look look at the hierarchical relations amongst the CKM matrix elements, we have considered the PDG \cite{pdg24} representation of the CKM matrix, wherein, unitarity is built-in. In the PDG representation, within the level of fraction of a percent, the mixing angles capture the hierarchy of the well measured 3 CKM matrix elements. Further, the CP violating phase $\delta$ is almost equal to one of the angles of the unitarity triangle. This representation is given by   
\begin{equation}
V_{CKM} = \bp
c_{12}c_{13} & s_{12}c_{13} & s_{13}e^{-i\delta} \\ 
-s_{12}c_{23}-c_{12}s_{23}s_{13}e^{i\delta} & c_{12}c_{23}-s_{12}s_{23}s_{13}e^{i\delta} & s_{23}c_{13} \\ 
s_{12}s_{23}-c_{12}c_{23}s_{13}e^{i\delta} & -c_{12}s_{23}-s_{12}c_{23}s_{13}e^{i\delta} & c_{23}c_{13}
\ep , \label{eq.pdg}
\end{equation}
where $c_{ij}= \cos \theta_{ij}, ~ s_{ij}=\sin \theta_{ij}$ for i, j=1, 2, 3. The angles $\theta_{12}, ~\theta_{23}$, ~$\theta_{13}$ correspond to the 3 mixing angles and $\delta$ is the CP-violating phase.  The angles $\theta_{ij}$ can be chosen to lie in the first quadrant, so  $0\leq s_{ij},  c_{ij}\leq 1$. Further for this  representation,  to a very good approximation (less than a fraction of a percent), the CKM parameters  $V_{ub}$, $V_{us}$ and $V_{cb}$ can be considered equal to the sines of the three mixing angles, i.e., $s_{13}$, $s_{12}$ and $s_{23}$ respectively. 
Experimentally, it is well established that $s_{13}\ll s_{23}\ll s_{12}$  and for small angles $s_{ij}\ll c_{ij}$.
 
In order to rigorously arrive at hierarchical relations amongst the CKM matrix elements, which is somewhat non trivial, we first discuss two interesting rephasing invariant relations $A_1$ and $A_2$ between the squares of the elements of the CKM matrix, also referred to as aymmetries \cite{asy1, asy2}. These can be obtained from the normalization conditions of the CKM matrix which are expressed as
 \be|V_{ud}|^2+|V_{us}|^2+|V_{ub}|^2 = 1 
 ,\label{r1}\ee
 \be 
 |V_{cd}|^2+|V_{cs}|^2+|V_{cb}|^2 = 1 
 ,\label{r2}\ee
 \be 
 |V_{td}|^2+|V_{ts}|^2+|V_{tb}|^2 = 1 
 ,\label{r3}\ee
 \be 
 |V_{ud}|^2+|V_{cd}|^2+|V_{td}|^2 = 1 
 ,\label{c1}\ee
 \be 
 |V_{us}|^2+|V_{cs}|^2+|V_{ts}|^2 = 1 
 ,\label{c2}\ee
 \be 
 |V_{ub}|^2+|V_{cb}|^2+|V_{tb}|^2 = 1 
 .\label{c3}\ee

The relation $A_1$ corresponds to the relations involving off diagonal elements about the main diagonal of the CKM matrix, i.e., $V_{ud}-V_{cs}-V_{tb}$ while the relation $A_2$ refers to the one involving off diagonal elements about the anti-diagonal of the CKM matrix, i.e.,  $V_{ub}-V_{cs}-V_{td}$. 
Using equations (\ref{r1}) and (\ref{c1}), one gets
 \be 
 |V_{us}|^2-|V_{cd}|^2=|V_{td}|^2-|V_{ub}|^2, \ee 
 similarly, from equations (\ref{r2}), (\ref{c2}) and from (\ref{r3}), (\ref{c3})
one obtains respectively 
 \be 
 |V_{us}|^2-|V_{cd}|^2=|V_{cb}|^2-|V_{ts}|^2 ~~~~~\text{and} ~~~~~|V_{td}|^2-|V_{ub}|^2=|V_{cb}|^2-|V_{ts}|^2 . \ee
  The above three relations are referred to as asymmetry $A_1$   given by 
 \be A_1: ~~~~~~ |V_{us}|^2-|V_{cd}|^2~~=~~|V_{td}|^2-|V_{ub}|^2~~ =~~|V_{cb}|^2-|V_{ts}|^2. \label{a1}\ee
 
Further, from equations (\ref{r1}) and (\ref{c3}), one obtains 
 \be 
 |V_{us}|^2-|V_{cb}|^2=|V_{tb}|^2-|V_{ud}|^2, \ee
  using equations (\ref{r2}), (\ref{c2}) as well as (\ref{r3}), (\ref{c1}), one obtains respectively 
 \be 
 |V_{us}|^2-|V_{cb}|^2=|V_{cd}|^2-|V_{ts}|^2 ~~~~~\text{and}~~~~~  |V_{tb}|^2-|V_{ud}|^2=|V_{cd}|^2-|V_{ts}|^2 .\ee
  These above relations correspond to asymmetry $A_2$ given by   \be A_2:~~~~~~ |V_{us}|^2-|V_{cb}|^2~~=~~|V_{tb}|^2-|V_{ud}|^2~~ =~~|V_{cd}|^2-|V_{ts}|^2. \label{a2}\ee

In order to arrive at the hierarchy among the CKM matrix elements,   we first consider the asymmetry $A_1$, mentioned in equation (\ref{a1}). Using PDG parametrization, the first difference of $A_{1}$  becomes 
 \be \begin{split}|V_{us}|^2-|V_{cd}|^2&= (s_{12}c_{13})^2-(-s_{12}c_{23}-c_{12}s_{23}s_{13}e^{i\delta})^2\\&
 = s_{23}^2(c_{13}^2-c_{12}^2)- s_{12}^2c_{23}^2s_{13}^2-2s_{12}c_{12}s_{23}c_{23}s_{13}\cos \delta. \end{split} \ee
 Noting that $s_{13}\ll s_{12} ~\text{and}~ s_{23}$, the above expression reduces to   
\be |V_{us}|^2-|V_{cd}|^2= s_{23}^2(c_{13}^2-c_{12}^2).\ee Similarly the other two differences also yield the same, implying \be A_1:~~~ |V_{us}|^2-|V_{cd}|^2~~=~~|V_{td}|^2-|V_{ub}|^2 ~~=~~|V_{cb}|^2-|V_{ts}|^2~~=~~ s_{23}^2(c_{13}^2-c_{12}^2).\label{a1v} \ee
Noting that $c_{13}> c_{12}$, one gets \be |V_{us}|>|V_{cd}|, ~~~~~|V_{td}|>|V_{ub}|~~~~~ \text{and }~~~~~ |V_{cb}|>|V_{ts}|. \label{hira1}\ee
Further, considering asymmetry $ A_2$ and using PDG parametrization, one gets
\be   A_2:~~~|V_{us}|^2-|V_{cb}|^2~~=~~|V_{tb}|^2-|V_{ud}|^2 ~~=~~|V_{cd}|^2-|V_{ts}|^2~~=~~ c_{13}^2(s_{12}^2-s_{23}^2). \label{a2v}\ee
 Since  $s_{12}>s_{23}$, one gets
 \be |V_{us}|>|V_{cb}|,~~~~~|V_{tb}|>|V_{ud}|~~~~~\text{and}~~~~~|V_{cd}|>|V_{ts}|.\label{hira2}\ee

It may be noted that equations (\ref{hira1}) and (\ref{hira2}) denote 6 inequalities among 8 elements of the CKM matrix. In order to find hierarchy among the 9 elements of the matrix, we first need to arrive at few more inequalities or orderings among the elements.
To this end, we consider
 \be |V_{ud}|^2-|V_{cs}|^2=
 c_{12}^2(c_{13}^2-c_{23}^2),\ee the expression on the RHS of the above equation is positive, therefore it implies
  \be |V_{ud}|>|V_{cs}|.\ee Combining this inequality with the second expression of equation (\ref{hira2}), one gets
 \be |V_{tb}|> |V_{ud}|>|V_{cs}|. \label{tbudcs}\ee
 Further, 
\be|V_{cs}|^2-|V_{us}|^2= c_{12}^2c_{23}^2-c_{13}^2s_{12}^2     ,~~~~\text{for}~~ c_{ij}> s_{ij}, ~~\text{one gets}~~|V_{cs}|>|V_{us}|.\label{csus}\ee
On combining the above ineqality along with the  ordering mentioned in  equation (\ref{tbudcs}) leads to \be |V_{tb}|> |V_{ud}| >|V_{cs}| >|V_{us}| \label{tbudcsus}. \ee
 Again,
 \be |V_{cd}|^2-|V_{cb}|^2= c_{23}^2(s_{12}^2+c_{13}^2)-c_{13}^2, ~~~ \text{implies}~~~~~|V_{cd}|>|V_{cb}|.\label{cdcb}\ee 
 The above inequality as well as the first and last inequalities mentioned in equation (\ref{hira1})  yield
 \be |V_{us}|> |V_{cd}|>|V_{cb}|>|V_{ts}| \label{uscdcbts}. \ee
 On combining the orderings mentioned in equations (\ref{tbudcsus}) and (\ref{uscdcbts}) one gets
 \be |V_{tb}|> |V_{ud}|>|V_{cs}| >|V_{us}| > |V_{cd}|>|V_{cb}|>|V_{ts}|. \label{all7}\ee
 Using \be  |V_{ts}|^2-|V_{td}|^2= s_{23}^2(c_{12}^2-s_{12}^2),~~ \text{one gets}~~~~~ |V_{ts}|>|V_{td}|.\ee
 The above inequality along with the second expression of equation (\ref{hira1}) leads to \be |V_{ts}|> |V_{td}|>|V_{ub}| \label{tstdub}.\ee 
 Incorporating the above with the orderings mentioned in equation (\ref{all7}) one gets
 \be |V_{tb}|> |V_{ud}|>|V_{cs}| >|V_{us}| > |V_{cd}|>|V_{cb}|>|V_{ts}|> |V_{td}|>|V_{ub}| \label{all9}. \ee
The above hierarchy amongst the elements of the CKM matrix, arrived at rigorously using the constraints of unitarity and well known constraints on the 3 mixing angles, is in agreement with the one given in Ref. \cite{asy3}. Any observation leading to the violation of this would have important implications for the 3 generation Standard Model. 
  
 \section{Numerical construction of CKM matrix elements using unitarity}
 
To verify the above  mentioned hierarchy of the CKM matrix elements as well as to study other implications of unitarity,  we have first reconstructed the elements of the CKM matrix using constraints of unitarity. At present, we have several CKM parameters which are determined with good deal of accuracy, e.g., the matrix element $|V_{us}|=0.22431 \pm 0.00085$ is determined within an accuracy of fraction of a percent \cite{pdg24}. Also, there are several CKM parameters which are known within an error of few percent, e.g., the parameter $\sin2\beta$, representing angle $\beta$ of the unitarity triangle,  with its world average being $(22.6\pm0.4)^{\degree}$ \cite{pdg24, hflav23}. Similarly, the angle $\alpha$ of the unitarity triangle is also known within
a few percent level, e.g., the world average is $(84.1^{+4.5}_{-3.8})^{\degree} $ \cite{pdg24}.

 Considering PDG parametrization, using the well defined  parameters $|V_{us}|$, $|V_{cb}|$, $\alpha$ and $\beta$, we have made an attempt to find the  mixing angles $\theta_{12}$, $\theta_{23}$ and $\theta_{13}$ as well as  the CP violating phase $\delta$.
 For this  representation,  to a very good approximation (less than a fraction of a percent) we can consider the cosines of the three mixing angles, i.e., $c_{13}$, $c_{12}$ and $c_{23}$ to be equal to 1. Therefore, the CKM parameters  $V_{ub}$, $V_{us}$ and $V_{cb}$ can be considered equal to the sines of the three mixing angles, i.e., $s_{13}$, $s_{12}$ and $s_{23}$ respectively. Therefore, considering the PDG 2024 value \cite{pdg24} of $|V_{us}|$, one gets 
$$
 s_{12} \cong |V_{us}|=0.22431\pm 0.00085,
 ~\text{implying}~
 \theta_{12} = 0.2262 \pm 0.0009. $$
 Considering the value of $|V_{cb}|$ as advocated by Belle collaboration \cite{belle23}, we obtain
$$
 s_{23}\cong |V_{cb}|= (40.6\pm 0.9)\times 10^{-3}, ~ \text{implying}~
 \theta_{23} = 0.0406 \pm 0.0009. $$
 Keeping in mind deviation in the inclusive and exclusive determination of the element $V_{ub}$, in order to find $\theta_{13}$, we have evaluated $|V_{ub}|$ using a unitarity based analysis involving  the `db' triangle \cite{pap}.
 From this triangle, one gets
  \be
|V_{ub}|\equiv\frac{|V_{cb}||V_{us}|\sin\beta}{\sin\alpha}.\ee
Using the values of CKM matrix element $|V_{us}|$ and 
$|V_{cb}|$ mentioned above and the earlier mentioned values of $\alpha$ and $\beta$  as given by PDG 2024 \cite{pdg24}, 
 we get $$ s_{13} \cong |V_{ub}|=(3.5184\pm 0.1022)\times 10^{-3},\label{eq.ub}~\text{implying}~ \theta_{13}=   0.00352 \pm 0.00010. $$
This is a rigorous unitarity based value of $V_{ub}$, which is in agreement with values given in Refs. \cite{pap,nik}. This value of $V_{ub}$ implies the ratio $ \left|\frac{V_{ub}}{V_{cb}}\right|=0.0867\pm0.0032$,  in agreement with  measurements from $\Lambda_b\rightarrow\rho \mu\nu$ and $B_s\rightarrow K\mu \nu$ decays \cite{hflav23}. 

As a next step, we evaluate the phase $\delta$ corresponding to this representation, which is almost equal to the angle $\gamma$ of the unitarity triangle. This may be checked by expressing  the angle $\gamma$ of the unitarity  triangle  in terms of  the elements of the mixing matrix, i.e.,
\be
\gamma=arg\left[-\frac{V_{ud}{V^*}_{ub}}{V_{cd}{V^*}_{cb}}\right].\ee
This can be simplified further to obtain an expression of $\delta$ in terms of $\gamma$ as \cite{cartesian}
\be 
\delta= \gamma + \sin ^{-1}(\sin \gamma \frac{s_{13}c_{12}s_{23}}{s_{12}c_{23}}).
\ee
It can be easily checked numerically that $\delta$ and $\gamma$ differ only to the extent of fraction of a percent, implying $\delta \cong \gamma$. For the present unitarity based analysis,  $\gamma$ can be found using the closure property of the angles of the unitarity triangle yielding
\be
\delta \cong \gamma = (73.3 \pm 4.2)\degree. \ee
 
 After having found the three mixing angles and the phase $\delta$, we obtain the corresponding  numerical values of the CKM matrix elements, i.e.,
 \be V_{CKM}=\begin{pmatrix}
0.97451\pm0.00018 &0.22430\pm 0.00080 &0.00352\pm 0.00010\\
0.22416\pm0.00080&0.97371\pm0.00019&0.04060\pm0.00090\\
0.00876\pm0.00031&0.03980\pm0.00088& 0.99917\pm 0.00004
\end{pmatrix}\label{eq.ckm1}
 .\ee
A look at this matrix reveals that this shows an excellent overlap with the one obtained by PDG\cite{pdg24}
\be 
\begin{pmatrix}
0.97435\pm0.00016 &0.22501\pm 0.00068 &0.003732^{+0.000090}_{-0.000085}\\
0.22487\pm0.00068&0.97349\pm0.00016&0.04183^{+0.00079}_{-0.00069}\\
0.00858^{+0.00019}_{-0.00017}&0.04111^{+0.00077}_{-0.00068}& 0.999118^{+0.000029}_{-0.000034}
\end{pmatrix}. \ee
Using these unitarity based values of the matrix elements,  one can immediately find that the hierarchy among the CKM matrix elements as mentioned in equation (\ref{all9}) is clearly confirmed.
 
\section{Parameters J, $\epsilon_k$ and hierarchy of CKM matrix elements}

As a next step, we discuss equivalent uniatity ensured possibilities of parameters J and $\epsilon_k$ and their interpretation in terms of hierarchy of CKM matrix elements. The orthogonality conditions, mentioned in equation (2) lead to the Jarlskog rephasing invariant parameter J, defined in terms of the 3 mixing angles and 1 CP violating phase, e.g., corresponding to the PDG representation, we get
\be J = s_{12}s_{23}s_{13}c_{12}c_{23}c_{13}^2 sin \delta. \ee
For the PDG representation as well as 8 other parametrizations recently \cite{cartesian} constructed, the value of parameter J comes out to be $(2.99 \pm{+0.13})\times 10^{-5}$.  As expected, this value is same for all the representations, also largely being in agreement with the PDG value \cite{pdg24}, i.e., $(3.12^{+0.13}_{-0.12})\times 10^{-5}$.

In a recent work \cite{asy3} it has been shown that while evaluating J, expressed in terms of magnitudes of the CKM matrix elements, there emerge 126 unitarity ensured equivalent possibilities to consider 4 moduli of the mixing matrix elements. Interestingly, most of these are not able to reproduce the value of the parameter J and only 8 viable choices remain. It has been shown that all of these eight patterns include the two smallest CKM matrix elements $V_{ub}$ and $V_{td}$. Besides these two elements, one of the CKM matrix element is either $V_{cb}$ or $V_{ts}$, these being next in hierarchical order as mentioned in equation (\ref{all9}). In any of the eight choices, the fourth CKM matrix element can be either $V_{us}$/ $V_{cd}$ or $V_{ud}$/ $V_{cs}$, these being higher in hierarchy to the above mentioned elements. Interestingly, from these observations one may conclude that understanding the evaluation of parameter J is facilitated by the hierarchy of the CKM matrix elements.

The above analysis also brings to fore that while carrying out CKM phenomenology it is important to keep in mind that hierarchical nature of CKM matrix requires one to be careful while considering various equivalent possibilities for carrying out the analysis. To emphasize this further, we present here another example wherein it is clearly seen that not all of the 9 possible parametrizations of the CKM matrix are equally suitable to carry out particular phenomenological analysis. In principle, all the 9 representations of the CKM matrix are equivalent as far as their physical implications are concerned, however, these could differ in respect to their suitability for particular applications. In particular, while calculating certain CKM parameters involving approximations, it is likely that some representations may be more useful for understanding a particular phenomenon. 

To this end, while evaluating $\epsilon_k$, the CP violation defining parameter in  the $K-\bar{K}$ system, using the 9 possible parametrizations of the CKM matrix, one gets interesting results. We use the following standard expression, involving short distance effects dominated by `t' quark, to evaluate  $\epsilon_k$, i.e.,
\be\epsilon_k=  \frac{{G_F}^2{F_k}^2 \hat{B_k} m_k{M_W}^2\kappa_{\epsilon}e^{\iota \phi_{\epsilon}}}{12 \sqrt{2}\pi^2 \Delta m_k} Im [{{\lambda^*}_c}^2 \eta_{cc} S_0(x_c)+{{\lambda^*}_t}^2 \eta_{tt} S_0(x_t)+ 2{{\lambda^*}_c}{{\lambda^*}_t} \eta_{ct} S_0(x_c,x_t)],\label{eq.e}\ee
where ${\lambda^*}_i=V_{id}{V^*}_{is}$  for i=c,t, 
$\kappa_{\epsilon}$ is the correction to $\epsilon_k$ due to long distance effects with value $\approx 0.94\pm 0.02$ \cite{pdg24}. Also, $\hat{B_k}$ = $0.717\pm0.024$ \cite{pdg24} is the bag parameter determined from lattice QCD. $\phi_{\epsilon}\approx 43.5^{\degree} $ \cite{pdg24} and $S_0$ are the Inami-Lim functions \cite{il} defined as 
 \begin{equation}
\begin{array}{l}
\vspace{0.2cm}
S_0 (x_i)=\frac{4x_i-11x_i^2+x_i^3}{4(1-x_i)^2}-\frac{3 x_i^3 ln x_i}{2 (1-x_i)^3}, \\ 
S_0(x_i,x_j)=x_i x_j\left[\frac{ln x_j}{x_j -x_i}\left(\frac{1}{4}+\frac{3}{2(1-x_j)}-\frac{3}{4(1-x_j)^2}\right)+\frac{ln x_i}{x_i -x_j}\left(\frac{1}{4}+\frac{3}{2(1-x_i)}-\frac{3}{4(1-x_i)^2}\right)-\frac{3}{4(1-x_i)(1-x_j)}\right],
\end{array}
\end{equation}
 where $x_i=\frac{m^2_i}{m^2_W}$ \cite{s0} and $  \eta_{ij}$ are perturbative QCD corrections. The various other parameters used in calculation of $\epsilon_k$ are given in Table \ref{parameters}.
\begin{table}[h]
\centering
\caption{Inputs used for evaluating $|\epsilon_k|$. }
\vspace{0.5 cm}
\begin{tabular}{|c|c|c|}
\hline
Parameter & Value &Reference\\ \hline
$G_F$&$1.1663787(6) \times 10^{-5}~\rm{GeV}^{-2}$&\cite{2301.12375}\\
$m_k$&$497.611(13)~\rm{MeV}$&\cite{2301.12375}\\
$\Delta m_k$&$3.484(6)\times 10^{-12}~\rm{MeV}$&\cite{2301.12375}\\
$m_W$&$80.356(6)~\rm{GeV}$&\cite{2301.12375}\\
$\eta_{cc}$&$1.72(27)$&\cite{2301.12375}\\
$\eta_{ct}$&$0.496(47)$&\cite{2301.12375}\\
$\eta_{tt}$&$0.5765(65)$&\cite{2301.12375}\\
$F_k$&$155.7(3)~MeV$&\cite{2301.02649}\\
$m_t$&$162.83(67)~GeV$&\cite{2301.02649}\\
$m_c$&$1.279(13)~GeV$&\cite{2301.02649}\\
 \hline
\end{tabular}
\label{parameters}
 \end{table}
 
For the 9 different representations mentioned in Table 2,  we have evaluated the parameter $\epsilon_k$. Following Ref.~\cite{review} and using the Cartesian representation 1 \cite{cartesian}, this being equivalent to PDG representation,  expressing $V_{cs}$, $V_{cd}$, $V_{ts}$ and $V_{td}$ in terms of the corresponding mixing angles and $\delta$ as well as using the numerical values of these inputs, we get 
 \be \epsilon_k= (2.119 \pm 0.245)\times 10^{-3},\ee this being largely in agreement with the one given by PDG, i.e., $(2.228\pm0.011)\times 10^{-3}$. The same exercise has been carried out for the remaining parametrizations. In principle, all the representations, being equivalent, should be providing equally appropriate fit to parameter $\epsilon_k$. Intriguingly, out of the 9 representations,  we find that the representations 1, 2, 3, 4, 7 and  8  are able to provide an appropriate fit to the parameter $\epsilon_k$. The other 3 representations, i.e., 5, 6 and 9 are very much off the mark. 
 
This can be understood when one closely examines the expression for parameter  $\epsilon_k$, given in equation (\ref{eq.e}), and the assumptions behind its derivation. Crucial ingredients of the formula are Inami-Lim functions and the parameters $\eta_{ij}$ which characterize the QCD corrections in the short distance limit. It can be easily understood that in the short distance limit, the correction $\eta_{cc}$, $\eta_{tt}$ and $\eta_{ct}$, characterizing the loops involving the c and t quark, would play a dominant role in the absence of the incorporation of the long distance effects. When this fact is coupled with imaginary part of the product of the $\lambda$ factors, one can easily understand why in the representations 5, 6 and 9, one is not able to reproduce the parameter $\epsilon_k$. In particular, the CKM matrix elements enter expression for parameter  $\epsilon_k$ through the factor ${\lambda^*}_i=V_{id}{V^*}_{is}$  (for i=c,t), implying that the elements $V_{cd}$, $V_{cs}$, $V_{td}$ and $V_{ts}$ are involved. Interestingly, for the representations 1, 2, 3, 4, 7 and  8 which predict correct value of $\epsilon_k$, the CP violating phase $\delta$ is associated with the elements $V_{td}$ and/or $V_{ts}$. 
Whereas, for the representations 5, 6 and 9, for which one is not able to reproduce  parameter $\epsilon_k$, the elements $V_{td}$ and $V_{ts}$ are both real and do not involve the CP violating phase $\delta$.
Relating this observation with the earlier mentioned hierarchical relation given in equation (\ref{all9}) one may, perhaps, conclude that the value of $\epsilon_k$ comes out to be correct when the phase $\delta$ is associated with the hierarchically smaller elements. Interestingly, this situation is echoing similar conclusions regarding the evaluation of Jarskog's rephasing invariant parameter J through unitarity ensured equivalent possibilities \cite{asy3}.

   \begin{table}[h]
   \renewcommand{\arraystretch}{1.5} 
\setlength{\tabcolsep}{1.8pt}
\centering
\caption{Value of parameter $\epsilon_k$ for different representations of CKM matrix}
\vspace{0.5cm}
\begin{tabular}{|c|c|c|}
\hline
Representation & Mixing matrix & $\epsilon_k$\\ \hline
1 &  $\begin{pmatrix}
 c_1c_2&c_1s_2&s_1\\
 -s_1c_2s_3-s_2c_3e^{-i\delta}& -s_1s_2s_3+c_2c_3e^{-i\delta}&c_1s_3\\
 -s_1c_2c_3+s_2s_3e^{-i\delta}&-s_1s_2c_3-c_2s_3e^{-i\delta}&c_1c_3
\end{pmatrix} $& $(2.119 \pm 0.245)\times 10^{-3}$
\\ \hline

2 &  $\begin{pmatrix} 
c_1c_2&s_1&c_1s_2\\
-s_1c_2c_3+s_2s_3e^{-i\delta}&c_1c_3&-s_1s_2c_3-c_2s_3e^{-i\delta}\\
-s_1c_2s_3-s_2c_3e^{-i\delta}&c_1s_3&-s_1s_2s_3+c_2c_3e^{-i\delta}
\end{pmatrix}  $   &
  $(2.119 \pm 0.184)\times 10^{-3}$
\\ \hline

3 & $
 \begin{pmatrix}
c_1c_2&-s_1c_2c_3+s_2s_3e^{-i\delta}&-s_1c_2s_3-s_2c_3e^{-i\delta}\\
s_1&c_1c_3&c_1s_3\\
c_1s_2&-s_1s_2c_3-c_2s_3e^{-i\delta}&
 -s_1s_2s_3+c_2c_3e^{-i\delta}
\end{pmatrix} $  & $(2.399 \pm 0.113)\times 10^{-3}$
\\ \hline

4 &$\begin{pmatrix}
-s_1s_2s_3+c_2c_3e^{-i\delta}  &-s_1s_2c_3-c_2s_3e^{-i\delta}&c_1s_2\\
c_1s_3& c_1c_3& s_1\\
-s_1c_2s_3-s_2c_3e^{-i\delta}&-s_1c_2c_3+s_2s_3e^{-i\delta}&c_1c_2 \end{pmatrix} $ &$(2.399 \pm 0.113)\times 10^{-3}$
\\ \hline

5 & $ \begin{pmatrix}
    c_1c_2& -s_1c_2s_3-s_2c_3e^{-i\delta}&-s_1c_2c_3+s_2s_3e^{-i\delta}\\
    c_1s_2&-s_1s_2s_3+c_2c_3e^{-i\delta}&-s_1s_2c_3-c_2s_3e^{-i\delta}\\
    s_1&c_1s_3&c_1c_3\end{pmatrix} $&
   $(0.16004 \pm 0.08821)$
\\ \hline

6 &  $\begin{pmatrix}
  -s_1s_2s_3+c_2c_3e^{-i\delta}&c_1s_3&-s_1c_2s_3-s_2c_3e^{-i\delta}\\
  -s_1s_2c_3-c_2s_3e^{-i\delta}&c_1c_3&-s_1c_2c_3+s_2s_3e^{-i\delta}\\
  c_1s_2&s_1&c_1c_2 \end{pmatrix} $ & $(0.16004 \pm 0.08821)$
\\ \hline

7 & $ \begin{pmatrix}
  c_1&s_1c_2&s_1s_2\\
  -s_1c_3&c_1c_2c_3-s_2s_3 e^{-i\delta} & c_1s_2c_3+c_2s_3e^{-i\delta}\\
  s_1s_3& -c_1c_2s_3-s_2c_3e^{-i\delta} & -c_1s_2s_3+c_2c_3e^{-i\delta}\end{pmatrix} $&  $(2.119 \pm 0.184)\times 10^{-3}$
\\ \hline

8 & $\begin{pmatrix}
    c_1c_2c_3-s_2s_3e^{-i\delta}&s_1c_2&-c_1c_2s_3-s_2c_3e^{-i\delta}\\
    -s_1c_3&c_1&s_1s_3\\
    c_1s_2c_3+c_2s_3e^{-i\delta}&s_1s_2&-c_1s_2s_3+c_2c_3e^{-i\delta} \end{pmatrix} $&
     $(2.399 \pm 0.113)\times 10^{-3}$
\\ \hline

9 &$\begin{pmatrix}
-c_1s_2s_3+c_2c_3e^{-i\delta}&c_1s_2c_3+c_2s_3e^{-i\delta}&s_1s_2\\
-c_1c_2s_3-s_2c_3e^{-i\delta}&c_1c_2c_3-s_2s_3e^{-i\delta}&s_1c_2\\
s_1s_3&-s_1c_3&c_1 \end{pmatrix} $ & $(0.16004 \pm 0.08821)$
\\ \hline

\end{tabular}

\end{table}

\section{ Summary and Conclusions} 
Recently \cite{asy3}, keeping in mind the hierarchical relation among the CKM matrix elements, it has been shown that out of the large number of unitarity ensured equivalent possibilities to determine the parameter J through 4 moduli of the mixing matrix elements, only a few are able to reproduce the value of J. These considerations provide a motivation for a re-visit of the CKM phenomenology wherein the hierarchy of the CKM matrix elements may play a similar role for evaluation of CKM parameters involving unitarity ensured equivalent possibilities.

In the present paper, we have rigorously re-visited the hierarchical relations among the CKM matrix elements using PDG parametrization incorporating unitarity constraints. We have checked these relations numerically by reconstructing the magnitudes of the CKM matrix elements by using well known CKM parameters. Further, similar to the example of evaluation of J \cite{asy3}, we have explored another case of evaluation of the CP violating parameter $\epsilon_k$. Interestingly, we find that using the 9 possible unitarity ensured equivalent independent parametrizations of CKM matrix, again not all of these provide an appropriate fit to parameter $\epsilon_k$. These results bring to fore the fact that because of the hierarchical nature of the CKM matrix elements as well as due to lack of requisite precision of these, in some cases analyses involving unitarity ensured equivalent possibilities may not lead to same physical results for all choices.

\section*{Acknowledgements}
The authors would like to thank the Chairperson, Department of Physics, Panjab University, Chandigarh, for providing the facilities to work.

  \end{document}